\documentclass[runningheads]{llncs}
\usepackage[T1]{fontenc}
\usepackage{caption}
\usepackage{subcaption}
\usepackage{graphicx}
\usepackage{amssymb}
\usepackage{amsmath}
\usepackage{color}
\usepackage{float}
\usepackage[hyphens]{url}
\usepackage{booktabs}
\usepackage{array}
\usepackage{tabularx}
\usepackage{tikz}
\usetikzlibrary{positioning}
\usetikzlibrary{patterns}
\usepackage{pgfplots}
\usetikzlibrary{arrows.meta, positioning, shapes.geometric}
\usepackage{algorithm}
\usepackage{algpseudocode}
\usepackage[dvipsnames]{xcolor}
\usepackage{setspace}
\usepackage{pifont}

\newenvironment{packeditemize}{
\begin{list}{$\bullet$}{
\setlength{\labelwidth}{0pt}
\setlength{\itemsep}{2pt}
\setlength{\leftmargin}{\labelwidth}
\addtolength{\leftmargin}{\labelsep}
\setlength{\parindent}{0pt}
\setlength{\listparindent}{\parindent}
\setlength{\parsep}{1pt}
\setlength{\topsep}{1pt}}}{\end{list}}
\begin{document}
\title{Detecting Aimbot Cheaters in MOGs}
\titlerunning{Detecting Aimbotters}
%
\author{
Salman Shaikh \and
Tao Ni  \and
Marc Dacier
}
%
%
%
\authorrunning{Shaikh}
%
%
\institute{}
\maketitle              
%


\begin{abstract}
Multiplayer Online Games (MOGs) have become a multi-billion dollar industry in the entertainment sector. However, the presence of cheaters undermines the experience of honest players and devalues the effort of game developers, as it directly affects player retention, competitive integrity, the legitimacy and trustworthiness of a game, and most importantly the overall revenue streams. Among various cheating techniques, visual aimbots represent an emerging threat. They use computer vision models to detect opponents from client screen captures rather than accessing game memory, making them completely undetectable by commercial kernel-level anti-cheat solutions.

In this paper, we introduce PATCH, a novel proactive defense strategy that deploys adversarial patches as in-game honeytokens to mitigate the presence of visual aimbot cheaters. Our approach centers on deliberately triggering the cheater's object detection model, enabling either direct detection, or rendering the game unplayable for the cheater via patch flooding on their viewport. We evaluate our approach on various criteria; analyzing the effectiveness of different patch sizes, scalability of patches to different screen resolutions, efficacy against diverse visual aimbot cheat configurations and also explore various YOLO models to assess patch transferability. Evaluation on a custom Unreal Engine game demonstrates over $90\%$ detection rate in white-box scenarios for almost all patch-sizes, and reaches $60\%-90\%$ cross-model transferability with larger patches. We further validate our approach on Fortnite, a commercial MOG, demonstrating real-world applicability.

\end{abstract}
\section{Introduction}\label{Intro}
The video game industry has experienced a massive growth over the past years, expected to reach a revenue of over 500 billion USD by the end of 2026 \cite{statista_games_worldwide_revenue}. This technological evolution has enabled the rise of Multiplayer Online Games (MOGs) (e.g., Fortnite, Valorant, Call of Duty, PUBG, and many more), which alone represent almost half of the aforementioned revenue \cite{fortune2026onlinegaming}. While single-player games and local multiplayer games still exist and significantly contribute to the gaming industry, MOGs have become the dominant force in the market. Players can pursue a professional career where they can earn financial rewards by playing and competing in their favorite games. However, this rampant rise alongside its respective monetary rewards have attracted malicious actors seeking unfair advantages by using various cheating tools and techniques \cite{apexLegendsEsportHacking}.

The arms race between game developers and cheaters has been going on for decades, with cheaters continuously developing new techniques to bypass defenses and evade detection. Among the many cheating techniques, aimbots are one of the most prevalent types of client-side cheats used in shooter games. These aimbots allow attackers to automatically move their crosshair or aiming reticle to track opponents. Traditional aimbots access the game's memory to extract positions of other players \cite{anwar2023extracting,karkallis2025vicevasivevideogame,robles2008online}. Such unauthorized access or manipulation can be detected by memory monitoring (anti-cheat systems) deployed in modern MOGs. However, a new class of aimbots has recently emerged that completely bypasses these commercial defenses, namely, visual aimbots. 

Rather than accessing the game memory, visual aimbots rely on the information rendered on the client screen. A visual aimbot contains many components including capturing the information from the screen, which is then fed to computer vision models (e.g., YOLO, Faster R-CNN, etc.) to detect the positions of opponents, and finally the coordinates of those positions are input into the game. Since visual aimbots can operate entirely outside of the client machine, traditional anti-cheat systems are completely ineffective. Hence, to mitigate the presence of visual aimbots, in this paper, we propose a novel proactive defense strategy that deploys adversarial patches as honeytokens. Specifically, our approach first generates carefully crafted patches designed to trigger the cheater's object detection model, then deploys these patches as Heads-Up-Display (HUD) overlay elements on the game's client at different locations and times. If the visual aimbot targets these honeytokens, the server can infer the presence of a cheater and take appropriate action.

We extensively evaluate our approach, covering multiple dimensions: 5 patch sizes ($10\times10$, $15\times15$, $20\times20$, $25\times25$ and $30\times30$), 3 different display resolutions ($1440\times1080$, $1920\times1080$, $3440\times1440$), both, same-machine and external-machine visual aimbot configuration and 3 different YOLO architectures (YOLO11m as base, white-box scenario, YOLO 12m and and YOLO26m to assess transferability of our proposed solution in a black-box scenario). Our evaluation on a custom Unreal Engine game demonstrates excellent results, achieving over $90\%$ detection rate for most patch sizes. Furthermore, when assessing the patches scalability for the same game but in a black-box scenario where the attacker may be using a different model, our patches demonstrate $60\%-90\%$ cross-model transferability with larger sizes of patches. Notably, even the lower bound of 60\% detection rate is sufficient for reliable cheater identification, since we only need to observe a few shots at a couple of honeytokens to infer the usage of a visual aimbot. Finally, to show that our proposed approach could be applied to readily available commercial titles, we validate our strategy on Fortnite. We summarize our contributions as follows:

\begin{packeditemize}
    \item We introduce a novel proactive defense strategy that creates adversarial 
    patches to be shown to the client's machines, serving as honeytokens that fool the attacker's model into targeting these patches.
    \item We evaluate our approach on a fully controllable custom game built using 
    the Unreal Engine, systematically testing patch parameters (size, location) and execution settings (display resolution, same-machine vs 
    external capture setups).
    \item Encouraged by the very positive results, we move on to demonstrate the applicability of our defense on Fortnite, a popular real-world MOG, and discuss the implications and limitations of our approach.
\end{packeditemize}

\section{Background and Related Work}\label{BackRel}
\subsection{Aimbots in Video Games}
Aimbots are software programs that are widely used in First Person Shooter (FPS) games (or, by extension, any kind of shooter game). Aimbots enhance the cheater’s performance by automating specific actions that require speed and precision. When equipped with an aimbot, the attacker's client automatically moves the cursor on top of opponents and begins tracking their movements \cite{karkallis2025vicevasivevideogame,robles2008online}, without
needing the cheater to manually locate the opponent and move the aiming reticle at them and
then shoot at each frame. This automation reduces reaction time while also achieving a high
level of efficiency that supersedes the reaction time of a normal player, giving cheaters a clear and unfair advantage over honest players. The literature identifies two types of aimbots: memory-based aimbots and visual aimbots.
\begin{packeditemize}
    \item \textbf{Memory-Based Aimbots}\label{memaim}: These aimbots are the most common type of aimbots and work by directly reading and manipulating the game memory. In this case, the cheat program attaches itself to the game process using OS-dependent APIs. Following that, it locates important data structures in the game memory which may contain information such as enemy positions, screen coordinates, and view matrices. With this data, the aimbot performs the necessary calculations and injects the results as device input into the game \cite{turner2025aimbot,karkallis2025vicevasivevideogame}. It is important to note that cheat developers often pair these type of aimbots with triggerbots, which acts as a complimentary component that automatically fires at detected targets, providing a complete targeting and shooting system. 
    
    
    \item \textbf{Visual-Based Aimbots}\label{visaim}: The emergence of visual aimbots represents a shift in how attackers now cheat. Instead of accessing the game memory and potentially risking detection by game specific anti-cheat systems, attackers have started using visual aimbots to bypass the defense entirely. Rather than extracting data from memory, visual aimbots rely on the same screen images seen by the player willing to cheat \cite{Nugroho2023Image,Chenxin2024Invisibility}. However, this process begins well before the actual gameplay starts:
    \begin{enumerate}
        \item Attackers first train a computer vision model (specifically; object detection) \cite{Chenxin2024Invisibility}, by capturing gameplay footage, labeling opponents with bounding boxes, and using these annotated images to train or fine-tune pre-existing models.
        \item The trained model is integrated with a screen capturing script and runs alongside the game. This approach never accesses the game memory but just performs inference on the live screen data to identify opponents.
        \item The visual aimbot then makes bounding boxes, extracts the coordinates and screen-space positions of detected target classes.
        \item Finally, it uses driver-level input simulation to move the cursor, achieving the same effect as traditional aimbots without ever accessing the variables or structures within the client's memory.
    \end{enumerate}
    The advantage of visual aimbots is that they operate non-intrusively, never accessing the game process, so typical anti-cheat systems focusing on memory checks become useless. 
\end{packeditemize}

\subsection{Existing Detection and Defense Methods}\label{detmeths}
Defenses against aimbots have been widely studied. We categorize existing approaches into two classes: mitigation against traditional memory-based aimbots and recent proposed solutions against visual aimbots.

\subsubsection{Defenses against Memory-Based Aimbots:}\label{defense_traditional}
Modern \textit{commercial anti-cheat systems} operate at the kernel-level of the operating system, monitoring any kind of process that attempts to gain unauthorized game memory access. They employ signature-based cheat detection as many client-side cheats, adapted to different games, share similar base characteristics \cite{BattleEye,EasyAntiCheat,RiotVanguard}. While these anti-cheat systems are effective against traditional aimbots, they are fundamentally unable to detect visual aimbots that operate externally and do not access the game memory. On the other hand, academic research has primarily focused on \textit{offline detection through game-dependent statistical analysis}. Studies based on different MOGs analyze different types of metrics, such as, kill/death ratios, headshot ratios, mouse movement, trajectory patterns to distinguish cheaters using aimbots from honest players \cite{alayed2013behavioral,Maberry2016UsingAA,han2015online,witschel2020aim,Xiao2023Detection,yu2012statistical}. However, these methods are not only game-specific (hence, not generalizable) but can also be evaded by attackers who configure their aimbots to mimic human-like behavior \cite{Kanervisto2022GANAimbotsUM}.

\subsubsection{Defenses against Visual Aimbots:}\label{defense_visual}
Recent work has begun addressing the advent of visual aimbots through \textit{disruption-based approaches}. Matsuhira et al. \cite{Matsuhira2024EffectsOA} propose adding adversarial patches to in-game characters to prevent detection by visual aimbots. However, these patches are continuously visible and also alter the character design itself (a key aspect in MOGs). More importantly, their visibility allows cheaters to retrain their models to filter them out. Chenxin et al. \cite{Chenxin2024Invisibility} improve upon this by applying imperceptible adversarial perturbations directly into the game frames that are rendered on the attacker's client machine. This effectively prevents the attacker's visual aimbot from targeting other players, while remaining invisible to cheaters'. Nonetheless, since cheaters remain undetected, they can iteratively retrain their models to become robust against such perturbation. 

Unlike these disruption based approaches, our work employs a different strategy. Rather than degrading the performance of the cheater's visual aimbot, we use adversarial patches as honeytokens to proactively trigger the attacker's model, enabling identification of the cheaters, and in the worst case, causing their visual aimbot to make the game "unplayable".

\section{Threat Model}\label{threatmodel}

\subsection{Attacker Model}\label{attmodel}
We consider an attacker operating a visual aimbot as described in Section \S\ref{visaim}. The attacker's visual aimbot runs non-intrusively and fully externally from the game process, relying solely on rendered screen frames as seen by the cheater. We assume the attackers may deploy their visual aimbot in one of the following configurations:
\begin{packeditemize}
    \item \textbf{Same-Machine Setup:} In this setup, the cheater operates both the game and the visual aimbot on the same machine. This is the most common configuration that is used in practice, where the visual aimbot captures the game's screen output directly through software-based screen capture OS-dependent APIs. 
    \item \textbf{External-Machine Setup:} If the anti-cheat solution that is shipped with a game, prevents any kind of screen capture to run alongside it, then the previously described setup will fail. To overcome this problem, a more capable attacker will utilize an external capture setup, where the host machine's screen output (where the game runs), is displayed via \texttt{HDMI} to a dedicated secondary machine that executes the visual aimbot. The secondary machine then relays input back to the host. This approach not only evades potential anti-cheat detections but also decouples the visual aimbot's computational demands from the game resources. 
\end{packeditemize}
The attacker's primary goal is to deploy a successful visual aimbot that automatically detects and targets opponents.

\subsection{Defender Model} \label{defmodel}
The defender (game developer/server) has full control over the visual information rendered on any connected client's screen, and can thus strategically display adversarial patches as honeytokens. If the attacker's visual aimbot targets these patches, the server can infer that the attacker is using a visual aimbot, and take appropriate action. The confidence in the server's conclusion is directly proportional to the number of targeted honeytokens. While targeting a single patch could indeed occur by chance, repeatedly shooting at multiple patches deployed across different screen locations provides conclusive evidence of a cheater. We consider the following two scenarios:
\begin{packeditemize}
    \item \textbf{White-Box Scenario}: We assume that the defender knows and has access to the object detection model used by the cheater. In this case, the defender can leverage the cheater's model to generate honeytokens designed to trigger it. This scenario corresponds to realistic situations where visual aimbot models are publicly available, either for free or for sale \cite{RootKit_AIAimbot,SunOner_SunOneAimbot}. Since most cheaters lack the expertise to develop their own models and instead purchase or download pre-trained ones, defenders can obtain the same models from the same sources to generate respective honeytokens. This scenario is developed in Section \S\ref{method}.
    \item \textbf{Black-Box Scenario}: In a worst case scenario, the defender will not know the attacker's model. We must consider the transferability of our approach. In practice, this means that the defender generates honeytokens using a known surrogate model and deploys them against an attacker using a different model. The success of the generalizability of our patches is discussed in Section \S\ref{results}.
\end{packeditemize}

\section{Methodology}\label{method}

\begin{figure}[tbp]
\centering
\resizebox{0.9\textwidth}{!}{
\begin{tikzpicture}[
    node distance=1.0cm,
    block/.style={rectangle, draw, fill=blue!15, text width=3.5cm, text centered, rounded corners, minimum height=2.5cm, font=\Large},
    usage/.style={rectangle, draw, fill=orange!15, text width=3.5cm, text centered, rounded corners, minimum height=2.5cm, font=\Large},
    arrow/.style={-Stealth, thick},
    label/.style={font=\Large, text=gray}
]

\node[block, fill=DarkOrchid!30!white] (training) {Training\\Images \\ (\ref{trainingimages})};
\node[block, fill=ProcessBlue!30!white, right=of training] (pi) {Patch\\Design \\ (\ref{patchgen})};
\node[block, fill=ProcessBlue!30!white, right=of pi] (gradient) {Gradient-based\\Optimization \\(\ref{gbo})};
\node[block, fill=ProcessBlue!30!white, right=of gradient] (deployment) {Patch \\Deployment \\(\ref{dnd})};
\node[usage, right=of deployment] (use) {Possible \\Use-Cases \\ (\ref{detection})};

\draw[arrow] (training) -- (pi);
\draw[arrow] (pi) -- (gradient);
\draw[arrow] (gradient) -- (deployment);
\draw[arrow] (deployment) -- (use);

\end{tikzpicture}
}
\caption{Defense Workflow: From Patch Generation to its Usage}
\label{fig:workflow}
\end{figure}
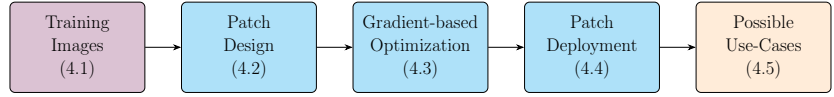

The goal of our approach is to create small adversarial patches 
that will be recognized by the attacker's object detection model as legitimate player targets. In other words, we aim to force the attacker's model to generate false positives, which can then be leveraged for detection or mitigation purposes. 

\subsection{Training Images}\label{trainingimages}
To optimize such patches as effectively as possible, we require a diverse training dataset that captures the visual context in which these patches will be deployed i.e. actual gameplay frames. Therefore, the initial step involves collecting a varied range of gameplay images that covers different scenes, elevations, backgrounds, characters, colors, etc. This diversity is important, to ensure that the patches develop intrinsic adversarial features that can remain effective regardless of the surrounding visual environment in which they are deployed.

\subsection{Patch Design}\label{patchgen}
Our approach exploits the inherent vulnerabilities of object detection models, specifically targeting their sensitivity to adversarial perturbations. 
Through gradient-based optimization, we can generate adversarial patches that, while appearing as arbitrary patterns to humans, contain the precise features required to trigger the model's detection logic. It is important to note that these adversarial patches are not random noise that happen to trigger the cheater's model by chance. Instead, they are deliberately optimized to maximize the model's classification score for the target class. The workflow involves:
\begin{packeditemize}
    \item Initializing a patch with random values as a starting point on top of gameplay images.
    \item Computing how the object detection model infers the images with these patches.
    \item Calculating gradients that indicate how to modify the patch pixels to increase detection confidence of the targeted class. 
    \item Iteratively updating the patch to make it detectable as the target class. 
\end{packeditemize}

The features that are contained in these patches could be heavily influenced by the patch size. Larger patches can hold more visual information and thus reliably trigger detections, however due to their larger size, they become more noticeable to any observers. On the other hand, smaller patches offer better stealthiness but may lack sufficient features to consistently trigger the model. Thus, this trade-off between stealthiness and size must be carefully balanced. Each patch is initialized as uniform random noise sampled from $\mathcal{U}(0.25, 0.75)$ independently for each pixel across all three RGB channels. This range of initialization values avoids extreme pixel values (i.e. near 0 or near 1), which could cause problems in computing appropriate gradients. This choice initializes a balanced starting point, allowing the optimization to occur in any direction to maximize model detection for the target class. Furthermore, to make sure that the generated patches are location-invariant, during the optimization, the patches are randomly positioned on training images at each iteration. This essentially prevents the patch from learning any location-specific information from the background images and ensures that it triggers a detection regardless of where it appears on the cheater's screen. 

\subsection{Gradient-based Optimization}\label{gbo}
The optimization process trains only the patch pixels while keeping the detector's weights entirely frozen. Specifically, we configure the patch as a trainable tensor $P \in \mathbb{R}^{h \times w \times 3}$ (where $h \times w$ is the patch size, and $3$ represents the RGB color channels), while all other parameters of the detector remain fixed and do not change throughout this process. 

The optimization objective is defined to maximize the detector's confidence that the patch represents the target class. Modern anchor-free object detectors output multiple predictions across the image. Each prediction consists of the respective bounding box coordinates $(x, y, w, h)$ and a class confidence score $S$ for the target class. To provide a broad gradient signal for optimization, we select the top $k=100$ predictions with the highest confidence scores across all spatial locations and compute: 

\begin{equation} 
\mathcal{L} = -\frac{1}{k} \sum_{i=1}^{k} S_{i} \label{losseq}
\end{equation} 

\noindent where $S_i$ is the raw class score of the $i$-th strongest prediction. By minimizing this negative mean (i.e., gradient ascent on confidence), we essentially force the patch to develop visual patterns that the detector strongly associates with the target class. The optimization process is described hereafter:
\begin{packeditemize}
    \item During each forward pass, the model processes training images with the initialized patch to generate detections.
    \item We then compute the partial derivatives $\frac{\partial \mathcal{L}}{\partial P}$ of the loss defined earlier \eqref{losseq}, with respect to each pixel in the patch. These partial derivatives are gradients that directs the optimization process to adjust pixel values to increase the model's detection confidence.
    \item An optimizer then updates the patch pixels based on these gradients: $P_{t+1} = \text{Optimizer}(P_t, \frac{\partial \mathcal{L}}{\partial P})$.
    \item Finally, the pixel values are clamped to the valid range of $[0,1]$ to maintain displayable colors.
\end{packeditemize}

\begin{figure}[tbp]
\centering
\resizebox{0.9\textwidth}{!}{
\begin{tikzpicture}[
    node distance=0.5cm,
    block/.style={rectangle, draw, fill=blue!15, text width=3.5cm, text centered, rounded corners, minimum height=4.0cm, font=\Large},
    decision/.style={diamond, draw, fill=orange!15, text width=3.2cm, text centered, aspect=1.0, inner sep=1pt, font=\Large},
    arrow/.style={-Stealth, thick},
    label/.style={font=\Large, text=gray}
]

\node[block, fill=green!15] (init) {Random Init\\$\mathcal{U}(0.25, 0.75)$\\per pixel, per channel};
\node[block, right=of init] (place) {Place Patch at Random \\ $(x,y)$\\on subset of images};
\node[block, right=of place] (forward) {Forward Pass\\through Frozen\\Detector};
\node[block, right=of forward] (loss) {Compute Loss\\$\mathcal{L} = -\frac{1}{k}\sum S_i$\\};
\node[block, right=of loss] (backprop) {Backpropagate\\Gradients to\\Patch Pixels};
\node[block, right=of backprop] (update) {Optimizer Update\\+ \\Clamp $[0,1]$};

\draw[arrow] (init) -- (place);
\draw[arrow] (place) -- (forward);
\draw[arrow] (forward) -- (loss);
\draw[arrow] (loss) -- (backprop);
\draw[arrow] (backprop) -- (update);

\draw[arrow, dashed] (update.south) -- ++(0,-0.8) -| (place.south) node[pos=0.25, above, font=\Large] {Iterative optimization};

\node[block, fill=yellow!15, below=2.5cm of place] (best) {Select best patch state during training};
\node[decision, right=1.5cm of best] (check) {$\geq$ Threshold\\confidence?};
\node[block, fill=green!35, minimum height=1.5cm, right=1.8cm of check, yshift=1.0cm] (accept) {Accepted\\Patch};
\node[block, fill=red!35, minimum height=1.5cm, right=1.8cm of check, yshift=-1.0cm] (reject) {Discarded\\Patch};

\draw[arrow] (update.east) -- ++(0.5,0) |- ([yshift=1.0cm]best.north) -| (best.north);
\draw[arrow] (best) -- (check);
\draw[arrow] (check.east) -- ++(0.8,0) |- (accept.west) node[pos=0.5, above, font=\Large] {Yes};
\draw[arrow] (check.east) -- ++(0.8,0) |- (reject.west) node[pos=0.5, below, font=\Large] {No};

\draw[arrow, dashed] (reject.south) -- ++(0,-0.6) -| (init.south) node[pos=0.25, below, font=\Large] {Retry with new seed};
\end{tikzpicture}
}
\caption{Patch generation pipeline}
\label{patch-pipeline}
\end{figure}
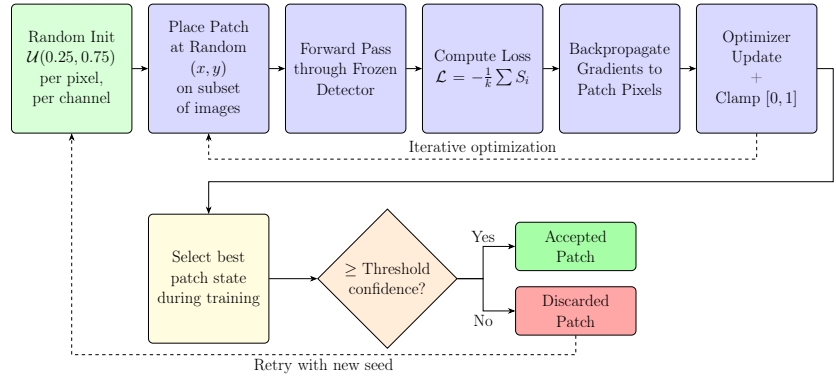

\noindent This process repeats over multiple iteration. Successful patches reaching sufficient confidence are selected as honeytokens, while failed attempts are discarded. The full pipeline is illustrated in Figure \ref{patch-pipeline}.

\subsection{Patch Deployment}\label{dnd}
Once the patches are generated, they must be deployed within the game environment.
It is crucial to deploy them in such a manner that they retain their adversarial properties. Our initial idea involved placing our patches as textures on in-game objects or making game objects out of patches. This method, however, had significant issues. Patches rendered through the game world's pipeline are subject to perspective transformations, lighting conditions, texture compression, distance-based adjustments, and much more. All of this variability, degrades the effectiveness of the carefully crafted patches. Preliminary tests quickly showed that it led to unreliable and extremely inconsistent results. In order to make sure our generated patches are displayed exactly as trained, we have pivoted to another approach. Instead of placing these patches in the game world, we display these patches directly as HUD (Heads-Up Display) overlay elements on the clients' screen. This method bypasses any transformations that the game world may have otherwise introduced, and allows the patches to retain their exact pixel values. This deployment strategy maintains the patch integrity, achieving both reliable detection and consistency, as shown in Section \S\ref{results}.

\subsection{Usage}\label{detection}
Once the patches are deployed and rendered on the attacker's screen, they cause the attacker's aimbot to generate a false positive as the patch is detected as a legitimate player target. Two distinct use cases can be considered for our patches depending on whether the attacker uses a visual aimbot and a triggerbot or solely operates a visual aimbot:

\begin{packeditemize}
    \item\textbf{Visual Aimbot + Triggerbot:} These systems will not only aim at the deployed patch but will also automatically fire at it, resulting in shots directed at the location of the patch. This gives the authoritative server reliable information of the presence of an attacker using a visual aimbot and can take appropriate action (such as banning the player from the match). 
    \item \textbf{Visual Aimbot \textit{Only}:} These systems will move the client's cursor at the patch, but no shot will be fired at it. In this case, the server does not receive any information. To address this scenario, the server can present various patches at different locations on the screen, essentially flooding their viewport and causing the attacker's model to continuously aim at different screen positions, rendering the game unplayable for the cheater. This approach, however, does not decrease the user experience of the honest players.
\end{packeditemize}


\section{Implementation}\label{implementation}

\subsection{Game Environment}\label{ge}
To assess the effectiveness of our method, we utilize the open source custom game built on the Unreal Engine, proposed by Shaikh et al. \cite{shaikh2025framework}. Using a custom game over which we have full control enables us to create a fully isolated evaluation environment while serving as a representative baseline for other Unreal Engine games. This is important because the Unreal Engine is one of the most widely used game engines in the industry; thus, our results can be easily generalized to a wide range of games. Due to the game's open-source nature, we are able to accurately simulate a real-world scenario, where an attacker runs their pre-trained visual aimbot model fully externally, while our defensive patches are deployed via in-game updates. We explain the complete process of developing the visual aimbot for this custom game from an attacker's perspective, followed by the generation of patches for the same game, from the defender's perspective. 

\subsection{Model Selection}\label{ms}
Initially, we must decide which model an attacker is likely to use. In the first step, we select YOLO11m as the base model that is used by the attacker, and therefore, the defender, to achieve their respective goals. In  a second step, we see how to transfer our results to scenarios where the attacker uses another method. 

YOLO11m consists of approximately 20 million parameters and requires 68.2 GFLOPs per forward pass. Such a model offers a balance between detection accuracy and inference speed, which is the most important criteria for real-time applications like a visual aimbot. Unlike two-stage detectors, YOLO processes the entire image in a single forward pass, significantly lowering inference latency. While this introduces a slight trade-off in accuracy compared to more complex models, a slight degradation of a few percentage points is completely acceptable as long as the model can still reliably and quickly detect other players. Although there are several object detection models in the literature, YOLO appears to be the most widely used model for real-time applications \cite{asmara2024yolo}. 

In this paper, we demonstrate the case study using YOLO11m, which is the medium-sized variant of the YOLO11 family. 
While smaller variants such as YOLO11n and YOLO11s (Nano and Small) offer higher frame rates, they do so at the cost of precision; conversely, larger variants like YOLO11l and YOLO11x (Large and Extra-Large) offer diminishing returns in accuracy (particularly due to the fact that, our case study has single-target class scenarios), while introducing more latency due to their larger magnitude of parameters. Furthermore, YOLO11m's architectural improvements allow it to outperform previous standards like YOLOv8m while utilizing fewer parameters \cite{Jocher_Ultralytics_YOLO_2023}. Another reason why we utilize YOLO11m as our primary model in this paper is because, it represents itself as one of the most stable standard for anchor-free detection. While older visual aimbots, such as the widely referenced RootKit AI-Aimbot \cite{RootKit_AIAimbot}, typically relied on anchor-based architectures (e.g., YOLOv5), with the development of newer and more efficient anchor-free models, the visual aimbot developers are shifting towards relying on these anchor-free models. Modern implementations like the SunOne Aimbot \cite{SunOner_SunOneAimbot} have adopted anchor-free models like YOLO11 for improved performance and lower latency. Therefore, selecting YOLO11m is both practical and consistent with what known visual aimbots currently employ. Newer families of YOLO such as YOLO12 and YOLO26 have also been recently introduced, and we show the scalability of our approach to newer versions in Section \S\ref{results}.

Rather than training the entire model from scratch, we use the YOLO11m model provided by Ultralytics \cite{Jocher_Ultralytics_YOLO_2023}, pre-trained on the COCO dataset. This approach allows us to leverage general visual features learned from a large dataset, such as edge detection, texture recognition, and shape primitives, etc, learned in the early layers. We then fine-tune this model with our own dataset. Only the final detection layer is modified to accommodate our single "player" target class, while the backbone remains unchanged. Although training from scratch might yield slightly better accuracy, fine-tuning is sufficient for our goal of achieving a fast and relatively accurate detection model for our visual aimbot. We could have experimented with various hyperparameters to develop an even better visual aimbot. However, the main goal is to show its efficiency and consistent reliability. Thus, we opted for the following parameters:
\begin{packeditemize}
    \item Epochs: We determined that 100 epochs is sufficient for the model to adapt its pre-trained weights to our custom gameplay dataset.
    \item Batch Size: A batch size of $8$ provides sufficient gradient noise to help the model distinguish the "player" class from complex background elements.
    \item Input Resolution: The input resolution was set to align with the standard training configuration of the YOLO model (640$\times$640), ensuring consistency with the pre-trained weights.
    \item Optimizer and Leaning Rate: Finally we used Adam, an adaptive gradient descent optimizer with a learning rate of $0.1$ for gradient-based optimization.
\end{packeditemize}

\subsection{Visual Aimbot and Patch Generation}\label{bothgen}
\begin{figure}[tbp]
    \centering
    \includegraphics[width=0.75\textwidth]{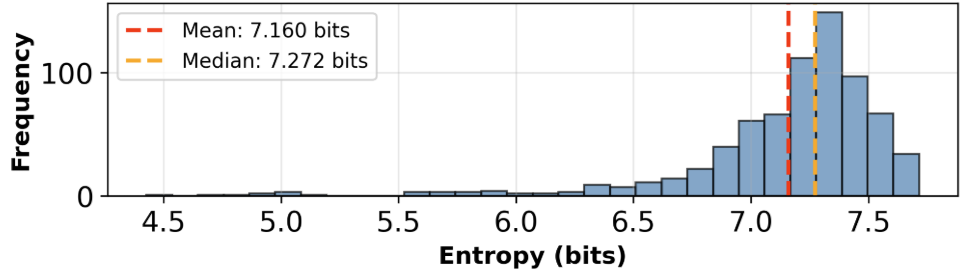}
    \caption{Entropy distribution of game data (8-bit scale)}
    \label{entropy_ds}
\end{figure}

To simulate a realistic attacker following the threat model described in Section \S\ref{threatmodel}, we construct a visual aimbot using the game environment and model configuration described in Sections \S\ref{ge} and \S\ref{ms}. Screen recordings of the custom game are captured in a game session and then split at a $1$ frame/second interval for manual annotation. Each extracted frame is carefully considered to ensure it contains distinct characteristics, such as featuring one or more characters, different lighting conditions, elevations, scales, etc. The manual annotation process focuses on identifying the coordinates of the opponent in varying conditions. For simplicity, we define the model's objective to focus and detect a single class target, "player". While an attacker could indeed train their model to detect different targets based on the game, a single class in this case is enough to demonstrate the mechanics and efficiency of a visual aimbot. Our annotated dataset (totaling 900 images) is used to train the model via 5-fold cross-validation, instead of a single train-validation split. This ensures correct performance evaluation across multiple independent validation sets. To quantitatively demonstrate this visual diversity in the training dataset, we computed the per-frame pixel-level entropy across the entire dataset. As shown in Figure \ref{entropy_ds}, the frames exhibit high entropy values (mean: 7.16 bits, median: 7.27 bits), which is almost the theoretical maximum of $8$-bits. The trained model's weights are subsequently used as the frozen target for patch generation, following the methodology explained in Section \S\ref{method}. Evaluation results for both the visual aimbot and the generated patches are presented in Section \S\ref{results}.

\section{Evaluation and Results}\label{results}
This section presents a comprehensive evaluation of our proposed solution. We begin by validating the visual aimbot's detection performance from an attacker's perspective to establish the baseline. We then evaluate our adversarial patches that are randomly rendered on the attacker's screen for 100ms, covering multiple avenues: different screen resolutions, capture configurations (same v.s. external machine capture), effectiveness of patch sizes, and their scalability to unseen models. All patch analysis experiments were conducted across multiple trials (5 patch sizes $\times$ 100 patches $\times$ 10 independent runs), and we report the average results to ensure statistical robustness. 
Finally, we demonstrate our approach's real world applicability through a proof-of-concept study on Fortnite. 

\subsection{Attacker's Visual Aimbot}
\begin{figure}[tb]
    \centering
    \includegraphics[width=0.75\linewidth]{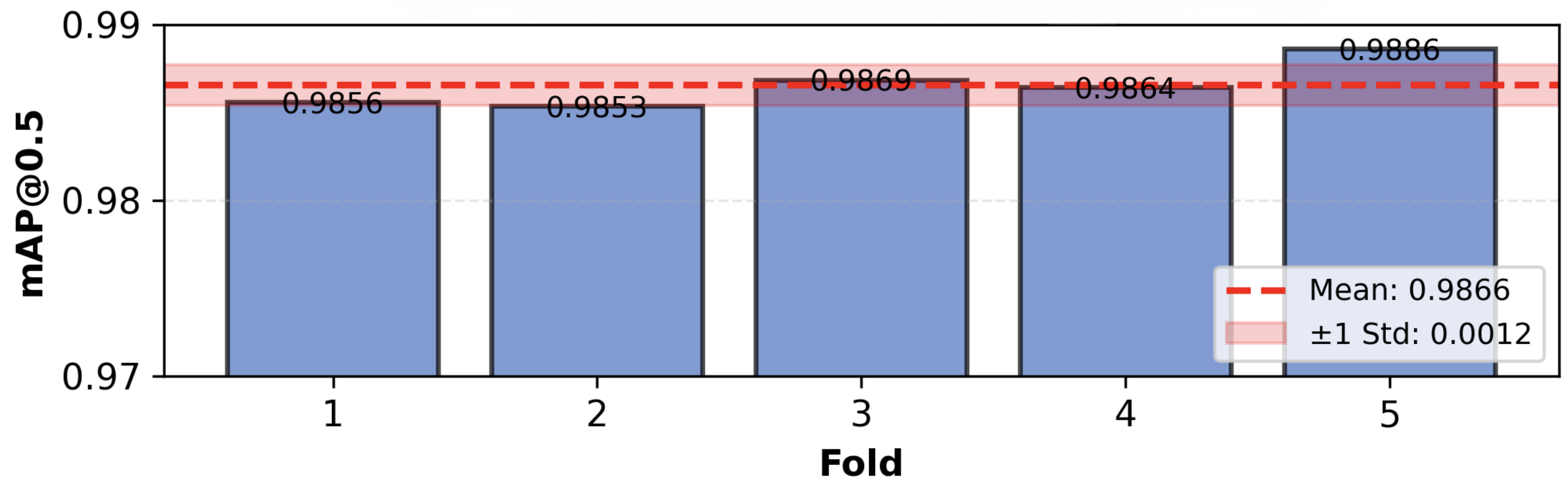}
    \caption{mAP@0.5 scores across 5-fold cross-validation}
    \label{kfold-map}
\end{figure}

\begin{figure}[tb]
    \centering
    \includegraphics[width=0.75\linewidth]{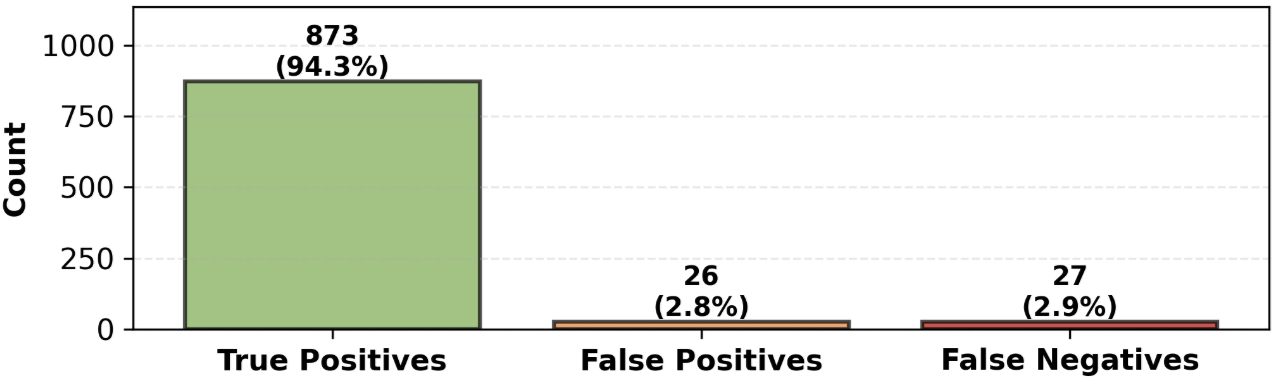}
    \caption{Aggregated TP, FP and FN across all validation folds}
    \label{kfold-confusion}
\end{figure}

We employed 5-fold cross-validation on our dataset of 900 images. In each fold, 4/5 of the data, i.e. 720 images were used for training and 1/5, i.e 180 images were used for validation, rotating through all folds to ensure each image was used for validation exactly once. Figure \ref{kfold-map} shows the mAP@0.5 scores across all five folds. The metric mAP@0.5 (mean Average Precision at $50\%$ IoU (Intersection over Union) threshold) measures how accurately the model detects players when predicted bounding boxes overlap ground-truth boxes by at least 50\%, ensuring that the predicted bounding boxes are neither too small nor too large. All folds achieve mAP@0.5 above 0.985, demonstrating consistently high detection accuracy with very little variation. The dashed red line shows the mean (0.986), and the shaded region represents $\pm1$ standard deviation (0.001). Furthermore, Figure \ref{kfold-confusion} shows the classification outcomes across all validation folds. A detection is considered a True Positive (TP) if the IoU $\geq$ 0.5 with the ground truth. The visual aimbot correctly detected 873 player instances (TPs, 94.3\%), while producing only 26 false detections on non-player objects (False Positives (FP), 2.8\%) and missing 27 actual players (False Negatives (FN), 2.9\%). This high TP rate, accompanied with low FP rate is important as it means that the model has learned salient features that correctly identifies a player and does not waste resources detecting background objects or other irrelevant game elements. The low FP and FN rate together demonstrate that the model achieves high precision and high recall. 

\begin{figure}[tb]
    \centering
    \includegraphics[width=0.75\textwidth]{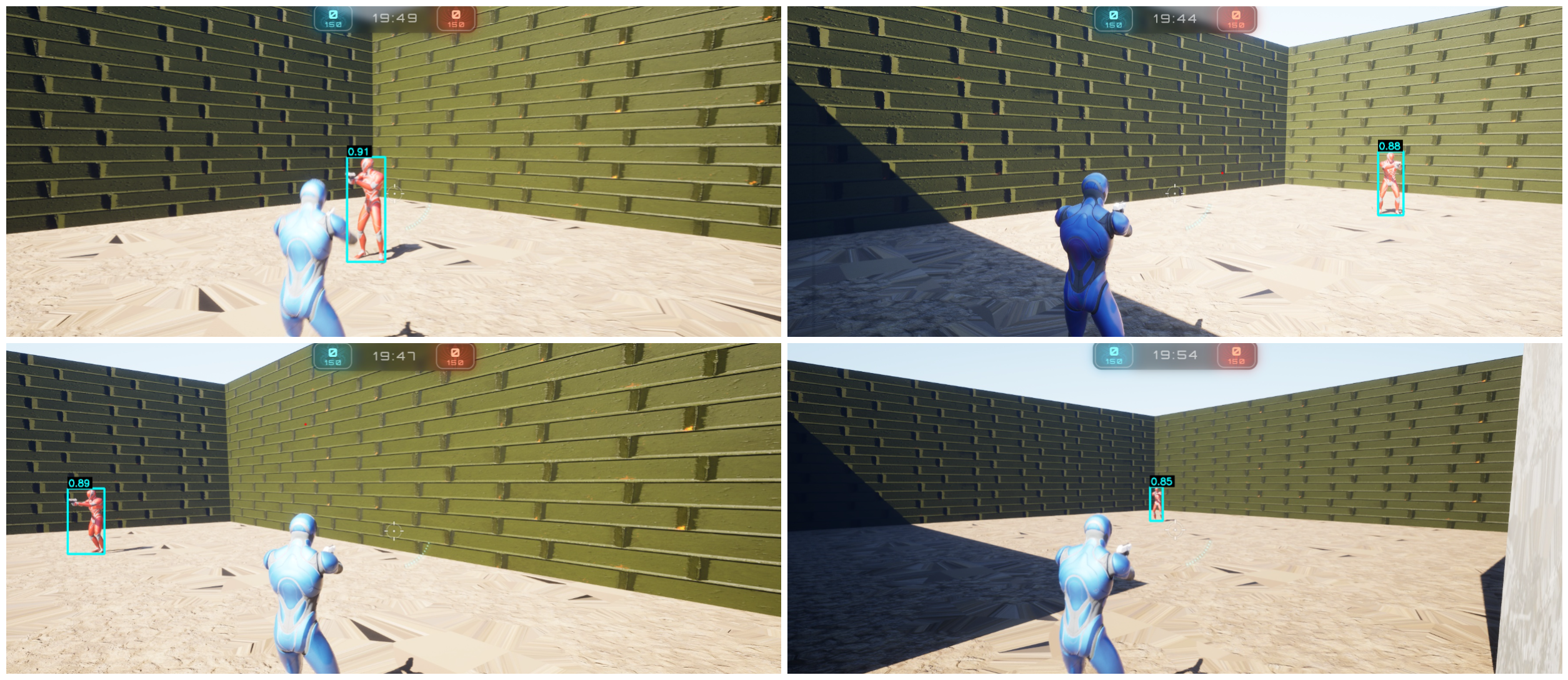}
    \caption{Real-time detection of opponents by the visual aimbot}
    \label{real_time}
\end{figure}

The final test involved running the visual aimbot alongside the game in a live scenario to verify that the model's strong validation performance is applicable in a practical scenario. We launched the visual aimbot, capturing the screen at $15$ FPS while the model performed inference on each frame. As shown in Figure \ref{real_time}, the visual aimbot successfully detects opponents across different screen positions with confidence scores consistently exceeding $0.80$. This confirms that a visual aimbot can be easily developed by attackers to gain unfair advantages with minimal problems, hence, the need for our proposed defense mechanism.

\subsection{Patch Test}
\begin{figure}[tbp]
    \centering
    \includegraphics[width=0.75\textwidth]{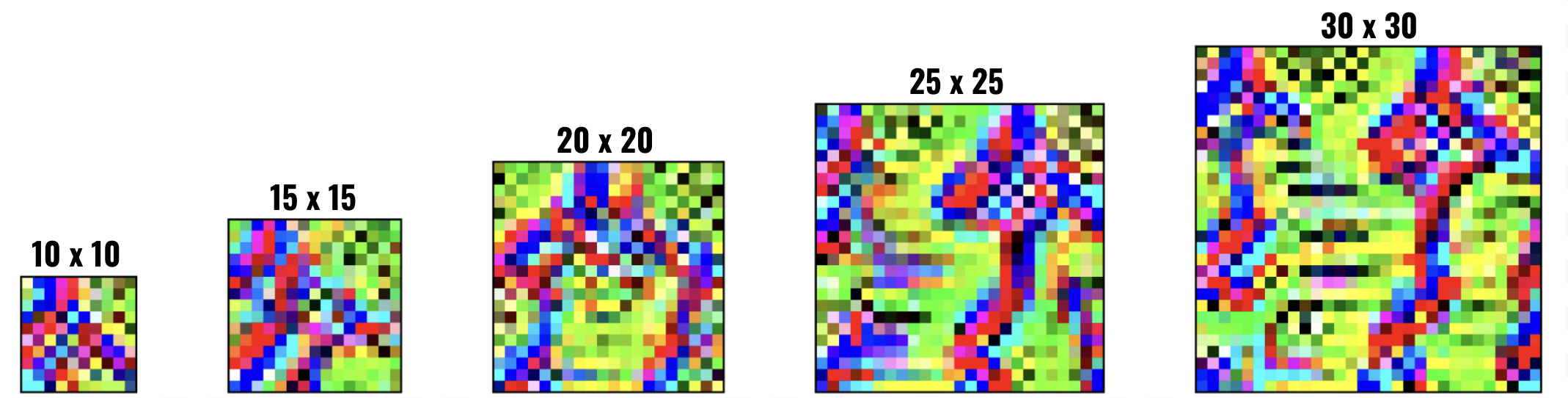}
    \caption{Final patches generated at different sizes}
    \label{patches}
\end{figure}

\begin{figure}[tbp]
    \centering
    \includegraphics[width=0.75\textwidth]{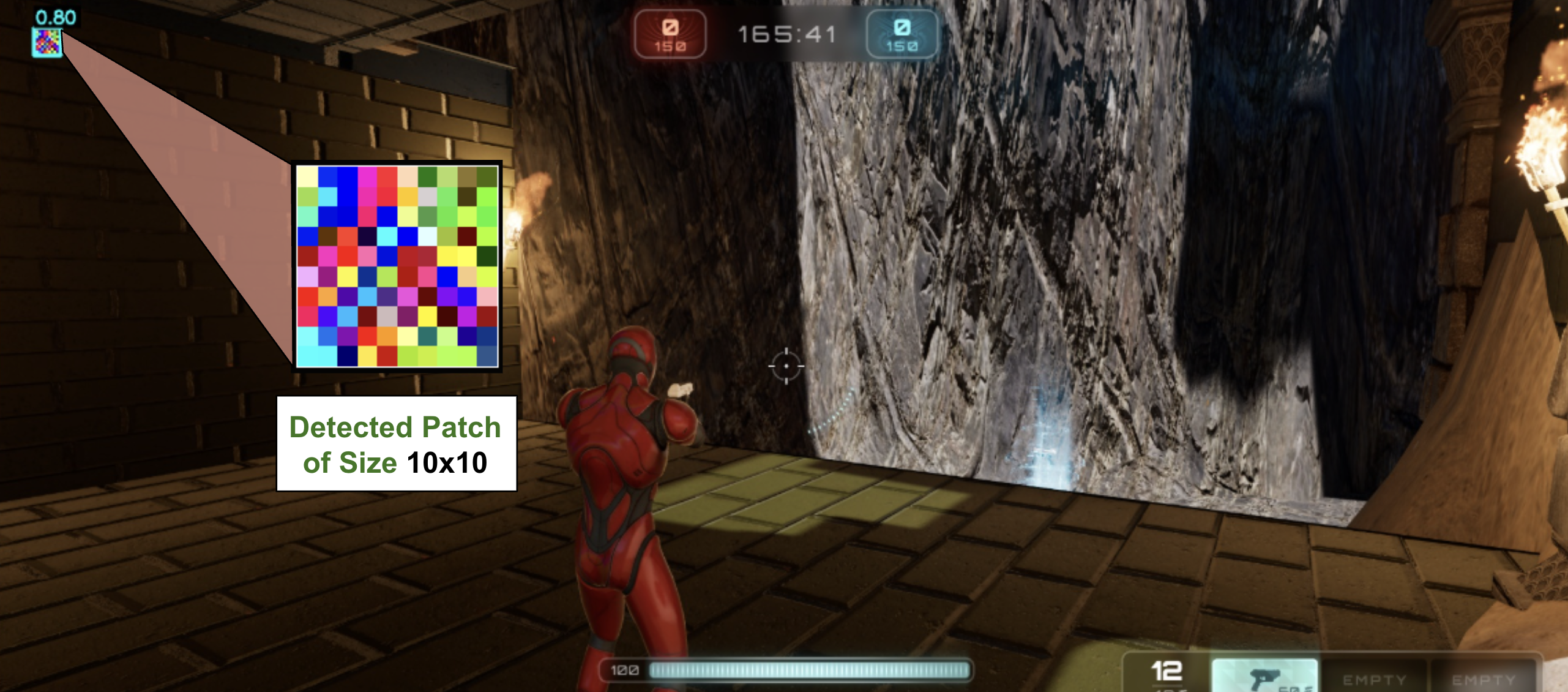}
    \caption{Example of a patch being deployed in-game}
    \label{patch_in_game}
\end{figure}

Following the methodology described in Section \S\ref{method}, we iteratively generated patches until we obtained $100$ patches per size ($5$ sizes, totaling 500 patches). Each patch was retained only if the detection confidence was $\ge 75\%$. Figure \ref{patches} shows an example of different patches of different sizes. We then tested the smallest patch ($10\times10$) in live gameplay to verify it would trigger the visual aimbot, shown in Figure \ref{patch_in_game}.

\subsection{Patch Effectiveness: Analysis}
To properly assess the effectiveness of our patches in the subsequent tests, our framework operates through two distinct logging mechanisms that together enable appropriate measurement of patch detection accuracy. On the defender's side (i.e. the game developer), when a patch is deployed on the attacker's screen, the game outputs a ground-truth log (in JSON format) recording the timestamp (UNIX time in milliseconds (ms)) and the absolute screen coordinates of the patch. Simultaneously, on the attacker's side, the visual aimbot independently generates its own detection log (also in JSON format) containing the timestamp and predicted bounding boxes for every detection event. Since we control both sides in our experimental setup, we can directly compare these two logs to determine detection accuracy. 
For each ground-truth patch deployment, we evaluate the potential corresponding detection predictions 
using a two-stage criterion. 
\begin{packeditemize}
    \item \textbf{Phase 1 - IoU-based Matching:} A detection is counted as a TP, if the predicted bounding box achieves an IoU $\ge 0.5$ with the ground truth patch region. This standard criterion ensures that we evaluate not only whether the patch triggered the model, but also whether the localization quality of the prediction is good.
    \item \textbf{Phase 2 - Fallback Step:} If the calculated IoU $< 0.5$, we apply a secondary check to account for cases where larger patches (e.g., $30\times30$) that indeed triggered the model, but produced a smaller bounding box within the actual patch region. In this case, a detection is still counted as TP if: (1) the center of the predicted bounding box falls within the ground-truth patch coordinates, and (2) the area of the predicted bounding box does not exceed $1.25\times$ the area of the ground-truth patch. This step enables us to handle scenarios where the model detects sub-regions of the patch rather than the entire patch, which we still consider a successful trigger. 
\end{packeditemize}
If both tests fail, the ground-truth patch is counted as a FN, indicating that the patch could not trigger the cheater's visual aimbot. Conversely, any detection that does not correspond to a ground-truth patch 
is counted as FP. 
This two-phase evaluation protocol ensures that we accurately measure the patch's ability to trigger the model and the spatial precision of the predicted bounding boxes.

\subsubsection{Screen Resolution Test:} An important preliminary evaluation focused on the robustness of our patches to different screen resolutions. The YOLO model, by default, expects an input resolution of $640 \times 640$ pixels, which is the resolution we used during training. However, real-world client displays vary widely. To maintain aspect ratio, YOLO applies letterboxing. It scales both dimensions of the captured frame by a uniform factor (derived from the longest dimension) to fit within $640 \times 640$, then adds padding to the shorter dimension to reach exactly $640 \times 640$. For example, consider a $1920 \times 1080$ frame. YOLO identifies the longest dimension (width = 1920) and computes the downscale factor as $1920/640 = 3$. Both dimensions are then scaled by this factor. The width becomes $1920/3 = 640$ pixels, and the height becomes $1080/3 = 360$ pixels. Finally, padding is added to the $640 \times 360$ scaled frame to reach the required $640 \times 640$ input size.

To ensure our patches maintain the same relative size in YOLO's input space after this transformation, we must render them at a correspondingly larger size on the original display. Specifically, a patch trained at $10 \times 10$ pixels (in the $640 \times 640$ training space) must be rendered as a $30 \times 30$ pixel texture on the $1920 \times 1080$ display to compensate for YOLO's 3$\times$ downscaling. Importantly, we do not modify the adversarial pattern itself. We simply render the patch texture at the scaled size using Unreal Engine's 2D HUD overlay system. The scaled patch dimensions are computed as:

\begin{equation}
    \text{scaled\_size} = \text{original\_size} \times \frac{\max(W_{\text{viewport}},\; H_{\text{viewport}})}{640}
\end{equation}

\begin{figure}[tbp]
    \centering
    \includegraphics[width=0.75\textwidth]{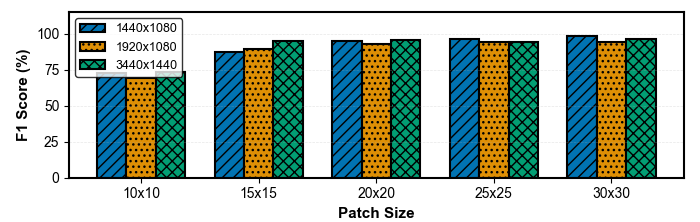}
    \caption{Scalability evaluation of patches across different display resolutions}
    \label{scalability}
\end{figure}

We evaluated patch performance across three different display resolutions: $1920 \times 1080$ (standard \cite{statcounter_screen_resolution_desktop_worldwide}), $1440 \times 1050$ (a smaller variation), and $3440 \times 1440$ (larger ultrawide format). The results in Figure \ref{scalability} show that the F1-scores of patches are consistently similar across different resolutions, confirming the scalability of our patches to different display sizes encountered in the real world. 

\subsubsection{Optimum Patch Size:} 
\begin{figure}[t]
    \centering
    \includegraphics[width=0.75\linewidth]{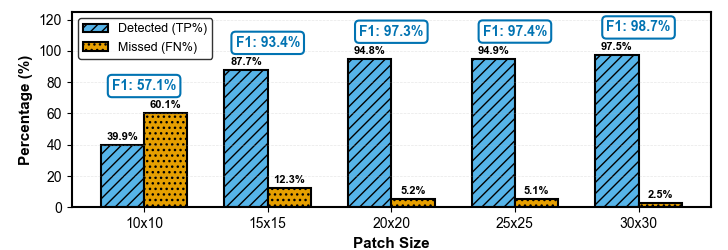}
    \caption{Comprehensive evaluation of patch detection across all sizes}
    \label{fulltest}
\end{figure}

The results shown in Figure \ref{fulltest} show that the detection rate of patches vary based on their size, as expected. The smallest patches of size 10$\times$10 achieved a F1-score of only 57.1\%, demonstrating the worst performance compared to other patch sizes. Interestingly, the patch size of 15$\times$15 achieved a significantly better F1-score of 93.4\%, while the remaining patches (20$\times$20, 25$\times$25, 30$\times$30) all achieved similar F1-scores between $97\%-99\%$. These results demonstrate a clear trade-off between patch sizes and their ability to trigger detections. While larger patches are more noticeable to attackers, they outperform smaller patches in triggering the visual aimbot. However, we can see that even a patch size of 15$\times$15 pixels reaches a detection rate of above 93\%, suggesting that this should be the minimum threshold size for reliable honeytoken deployment in real-world live game capture. 

\subsubsection{Same Machine Capture v.s. External Machine Capture:} 
\begin{figure}[tbp]
    \centering
    \includegraphics[width=0.75\linewidth]{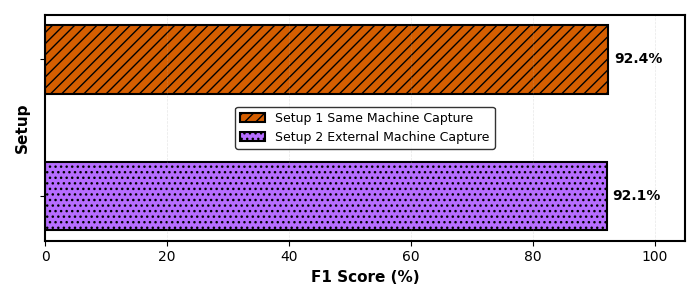}
    \caption{F1-score comparison between same and external-machine capture}
    \label{capture_comparison}
\end{figure}

To verify that patch effectiveness remains constant between different capture configurations, we evaluated all patches of size $15\times15$ under both, same-machine and external-machine setups. We selected this patch because it achieves a high F1-score while being more stealthier than the larger patch sizes. The patch was systematically displayed in the same sequence for both configurations, in order to maintain a controlled comparison. For the external setup, the screen output of where the game runs, is transmitted to a dedicated machine via a \texttt{CamLink} capture card, where the visual aimbot runs. Figure \ref{capture_comparison} shows that both configurations achieve equivalent F1-scores of approximately 92\%. The consistency between both setups validates the fact that game developers can reliably mitigate visual aimbots regardless of whether attackers use an external machine or not.

\subsubsection{Scalability to Newer Models of YOLO:}
\begin{figure}[tbp]
    \centering
    \includegraphics[width=0.75\linewidth]{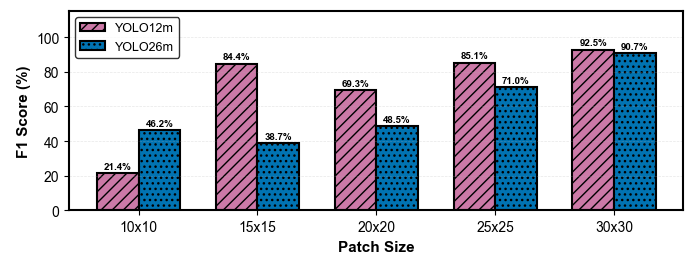}
    \caption{Comparison of the same patches across different YOLO versions}
    \label{recall_comparison}
\end{figure}

While our case study in this paper focused primarily on YOLO11m, the underlying principles of the way we generated and optimized the patches, are not specific to this model only. 
Instead of developing patches for distinct versions and demonstrating similar results we presented earlier, we wanted to check if the same patches we developed using YOLO11m, could potentially work for the newer versions of YOLO as well. This is crucial because if our patches work for YOLO12 and YOLO26 without any modifications, then this would show that our approach is not only effective for the current standard models, but also future-proof to some extent. To show this, we developed the visual aimbot from the attacker's perspective using YOLO12m and YOLO26m provided by Ultralytics. We then fed the same patches to these newer models and evaluated their performance in the same way. The results, shown in Figure \ref{recall_comparison}, confirm that the adversarial patches successfully transfer across the newer YOLO models. Nonetheless, their effectiveness varies significantly depending on both the target architecture and size of the patch itself. 

For the smallest patch size of $10\times10$, the F1-scores drop to approximately 21\% for YOLO12m and 46\% for YOLO26m, rendering this configuration almost impractical for cross-model transferability. For YOLO12m, patch performance exhibits an interesting pattern. We see that the $15\times15$ patch achieves about 84\% F1-score, while the $20\times20$ patch unexpectedly drops to 69\%, before recovering to 85\% at $25\times25$ and reaching the highest score of 92\% at $30\times30$. This dip at $20\times20$ could possibly be attributed to sub-optimal convergence during patch optimization for this particular size. YOLO26m, however, exhibits a more pronounced dependence on patch size, with F1-scores increasing generally monotonically from 48\% at $20\times20$ to about 71\% at $25\times25$ and reaching 90\% at $30\times30$. These findings are rather insightful. They indicate that while the adversarial patches do indeed generalize to newer YOLO architectures, their transferability is generally positively correlated with patch size, highlighting the importance of patch size selection in black-box scenarios (different models used by the defender and the attacker). 


\subsection{Real World Application: Fortnite}
To show that our methodology is not game dependent and that the same approach could be applied to any commercial games, we conduct a proof-of-concept test on Fortnite. We obtained an open-source visual aimbot provided by SunOne \cite{sunone_boosty}, which uses a YOLO12 model with publicly available \texttt{.pt} and \texttt{.onnx} weights. We captured 1,000 Fortnite gameplay frames to use as training images. We adopted a rigorous 80/20 train-test split protocol during the generation of the patches. Each patch was trained on 800 frames using gradient-based optimization (Section \S\ref{gbo}) with the frozen YOLO12 model, then evaluated exclusively on the 200 unseen test frames. However, we were unable to generate patches for the smallest patch sizes ($10\times10$, $15\times15$, and $20\times20$), which failed to meet the 75\% confidence threshold, despite multiple training attempts with different seeds. This limitation highlights the sensitivity of adversarial patch generation to many aspects; model parameters, patch size, game-specific visual characteristics, etc. Therefore, we report results only for the two larger patch sizes ($25\times25$, and $30\times30$) that successfully generated adversarial patches.

\begin{figure}[tb]
    \centering
    \includegraphics[width=1.0\linewidth]{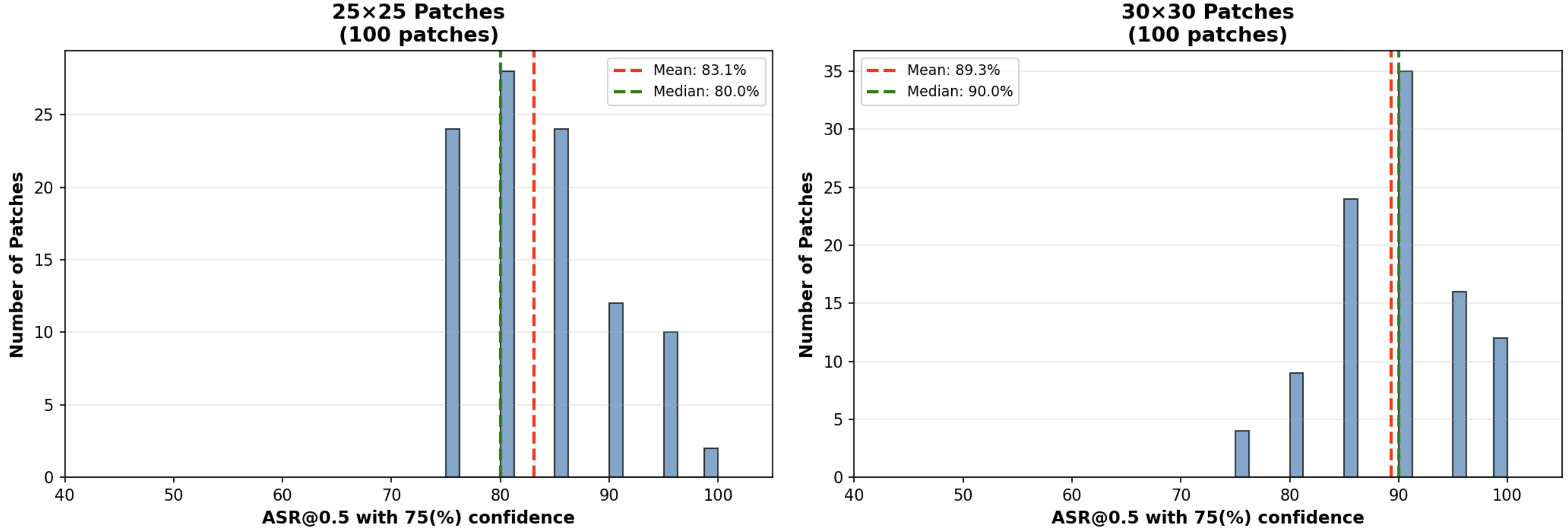}
    \caption{ASR@0.5 distribution: Patch sizes $25\times25$ (left), and $30\times30$ (right)}
    \label{fnquantitative}
\end{figure}

\begin{figure}[tb]
    \centering
    \includegraphics[width=0.70\linewidth]{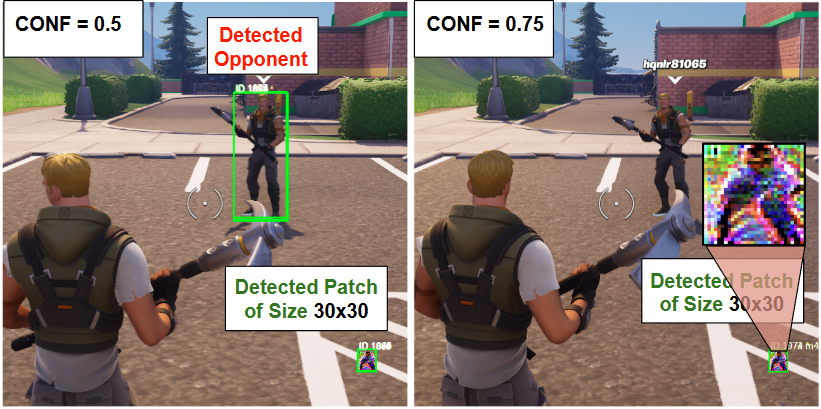}
    \caption{Detections at confidence 0.5 (left) and 0.75 (right)}
    \label{fortnite_confidence_comparison_comb}
\end{figure}

To evaluate how effectively the adversarial patches trigger detections, we measure the ability of each patch to produce correct detections on unseen test images. To quantify this, we use the Attack Success Rate at IoU 0.5 (ASR@0.5) with a confidence threshold of 75\%. This metric represents the percentage of test images in which the patch triggers a detection with an IoU of at least 0.5 with the ground truth, and where the detection confidence is at least 75\%. Figure \ref{fnquantitative} presents the ASR@0.5 distribution for both patch sizes ($25\times25$ and $30\times30$). The $25\times25$ patches achieve a mean ASR@0.5 of 83.1\% and a median of 80.0\%, with most patches falling between 75–85\% on unseen test frames. In comparison, the $30\times30$ patches perform better, reaching a mean of 89.3\% and a median of 90.0\%, with nearly all patches exceeding 80\% ASR@0.5.
For live-gameplay testing, patches are superimposed externally as overlays since we lack control over Fortnite's rendering pipeline. Notably, Figure \ref{fortnite_confidence_comparison_comb} demonstrates a critical finding. Our optimized $30\times30$ adversarial patch achieves higher detection confidence than legitimate opponents. At a 0.5 confidence threshold, both the opponent and patch are detected, but when the threshold is raised to 0.75, the opponent detection is suppressed while the patch remains detected. This confirms that our patches can trigger stronger model responses than real players. 


\section{Discussion}\label{discussion}

\subsection{Ethical Considerations}
This research was conducted entirely within controlled environments; a custom game over which we have full control, and isolated private matches in Fortnite where no other player was affected. The visual aimbot implementations used to illustrate an attacker's perspective and its evaluation, were developed solely for research purposes and have not been distributed. Similarly, no patches were deployed in public game servers to test their applicability.

\subsection{Limitations}
\subsubsection{Patch Effectiveness v.s. Stealthiness:} Our results revealed that, indeed, larger patches i.e. $30\times30$ consistently achieved highest F1-scores across all variants of YOLO, whether it was a white-box approach, or a black-box approach. However, this patch size is potentially noticeable by an attacker. On the other hand, smaller patch sizes i.e. $10\times10$ achieved suboptimal performance in both white-box and black-box scenarios. Hence, game developers must carefully balance these factors on their specific game dependent deployment scenarios

\subsubsection{Cross-Model Transferability:} Although our patches demonstrate reasonable transferability to YOLO12m and YOLO26m, performance degradation is evident, particularly for smaller patch sizes. Nonetheless, by deploying simultaneously, patches built for distinct models, game developers could potentially identify which specific model variant the cheater employs. This intelligence could enable adaptive countermeasures or inform threat analysis about prevalent visual aimbots. Extending our idea beyond YOLO variants to other object detection architectures (e.g., transformer-based) would help assess their generality and robustness across fundamentally different model designs.

\subsection{Future Work}
\subsubsection{Game-Based Patch Initialization:}
Our current approach initializes patches with uniform random noise sampled from $\mathcal{U}(0.25, 0.75)$. Future work could improve this patch initialization by computing common pixel block statistics depending on the game environment. This includes analyzing textures, lighting conditions, color palettes of maps, etc. Patches initialized with environment-specific sampling would blend more naturally into the game, allowing us to potentially gain better patch stealthiness and detection rates.

\subsubsection{Adaptive Mitigation Strategies:} Our defense strategy allows mitigation for attackers using a visual aimbot and a trigger bot (ban them) or if they are using solely a visual aimbot (flood their viewport with patches), as described in Section \S\ref{detection}. 
To extend coverage of more stealthy attackers, future approaches could rely on game-specific behavioral analysis systems to implement adaptive patch flooding algorithms. Players who are flagged as suspicious by other anti-cheat metrics (e.g., unusual movement patterns, camera rotations, etc.) could receive increased patch exposure, compared to other possibly honest players.


\section{Conclusion}\label{conclusion}
This paper presents the first proactive defense mechanism that uses adversarial patches as honeytokens to mitigate visual aimbots. The patches exhibit robust scalability across different display resolutions and maintain effectiveness even when displayed for only $100$ms. Our comprehensive evaluation demonstrates the effectiveness in both white-box and black-box scenarios. When the attacker's model is known, the adversarial patches achieve over $90\%$ F1-scores for almost all patch sizes. When the attacker's model is unknown, larger patches still transfer effectively achieving F1-scores between $60\%-90\%$. Furthermore, our POC on Fortnite, using an open-source visual aimbot, confirms that our methodology is applicable to popular commercial games. 

As visual aimbots become more prevalent, proactive detection mechanisms become increasingly important to maintain the integrity of MOGs. Our defensive strategy of using a honeytoken-based approach offers game developers a new tool, that transforms the attacker's own weapon, against themselves.

\begin{credits}

\subsubsection{\discintname}
The authors have no competing interests to declare that are relevant to the content of this article.
\end{credits}
%
%
%
\bibliographystyle{splncs04}
\bibliography{main}
%




\end{document}